# Hand-held 3D Photoacoustic Imager with GPS

Daohuai Jiang[1,2,3,†], Hongbo Chen[1,2,3,†], Yuting Shen[1], Yifan Zhang[1], Feng Gao[1], Rui Zheng[1,*], and Fei Gao[1,*]

[1] *Shanghai Engineering Research Center of Intelligent Vision and Imaging, School of Information Science and Technology, ShanghaiTech University, Shanghai 201210, China*
[2] *Chinese Academy of Sciences, Shanghai Institute of Microsystem and Information Technology, Shanghai 200050, China*
[3] *University of Chinese Academy of Sciences, Beijing 100049, China*
[†]*equal contribution*
* *zhengrui@shanghaitech.edu.cn; gaofei@shanghaitech.edu.cn*

**Abstract:** As an emerging medical diagnostic technology, photoacoustic imaging has been implemented for both preclinical and clinical applications. For clinical convenience, a handheld free-scan photoacoustic tomography (PAT) system providing 3D imaging capability is essentially needed, which has potential for surgical navigation and disease diagnosis. In this paper, we proposed a free-scan 3D PAT (fsPAT) system based on a hand-held linear-array ultrasound probe. A global positioning system (GPS) is applied for ultrasound probe's coordinate acquisition. The proposed fsPAT can simultaneously realize real-time 2D imaging, and large field-of-view 3D volumetric imaging, which is reconstructed from the multiple 2D images with coordinate information acquired by the GPS. To form a high-quality 3D image, a dedicated space transformation method and reconstruction algorithm are used and validated by the proposed system. Both simulation and experimental studies have been performed to prove the feasibility of the proposed fsPAT. To explore its clinical potential, *in vivo* 3D imaging of human wrist vessels is also conducted, showing clear subcutaneous vessel network with high image contrast.

## 1. Introduction

In the past two decades, photoacoustic (PA) imaging (also called optoacoustic imaging) has been widely applied for biomedical applications. PA imaging is a kind of hybrid imaging modalities that combines spectroscopic optical absorption contrast and deep acoustic penetration [1, 2]. The basic principle of photoacoustic imaging includes laser excitation, optical absorption and temperature elevation, thermoelastic expansion, PA wave emission, signals detection and image reconstruction [3, 4]. Generally, the implementation of PA imaging can be categorized as three modalities: photoacoustic microscopy (PAM), photoacoustic tomography (PAT) and photoacoustic endoscopy [5]. The PAM system has high spatial resolution suitable for skin and capillary imaging. To further classify PAM, it can be categorized as optical-resolution PAM (OR-PAM) and acoustic-resolution PAM (AR-PAM) [6, 7]. OR-PAM can achieve micrometer-level resolution with optical focusing, but its imaging deep is limited within 1 mm, which is difficult to realize 3D imaging with deep penetration. On the other hand, AR-PAM has better penetration than OR-PAM due to deeper acoustic focusing. However, its imaging speed is limited by the mechanical scanning shown in Fig. 1(a). Recently, microelectromechanical system (MEMS) based mirror has been applied for improving the imaging speed [8, 9], which is still suffering quilted limited field-of-view. Therefore, the PAM system is normally used for small region of interest (ROI) imaging with slow imaging speed, which cannot satisfy the real-time and large ROI requirements in clinical applications [10].

On the other hand, PAT system can realize real-time imaging with much larger ROI than PAM system [11-13]. For hand-held PAT system, a linear-array ultrasound transducer (UT) is applied for PA wave detection as shown in Fig. 1(b). This PAT system can only reconstruct 2D image of the UT's receiving plane. To realize 3D imaging, two conventional approaches are: 1). Using custom-designed 3D UT, such as spherical array shown in Fig. 1(c) [14-17]; 2). Mechanical scanning of the linear-array UT [18, 19]. Regarding the first approach, although custom-designed spherical-array UT can easily achieve 3D PA imaging, it is limited to fixed ROI, and expensive nonstandard UT fabrication cost. Regarding the second approach, it suffers: 1). The mechanical scanning part includes a stepper motor and a fixed movement rail, which is clumsy and inflexible for clinical application; 2). The imaging area is dependent on the mechanical stepper motor with limited movement distance; 3). The UT detection track is also fixed and cannot dynamically adjust during the scanning. Therefore, a free-scan 3D PAT (fsPAT) system based on conventional linear-array UT is highly needed to address these issues, which can realize flexible handheld scanning with large ROI in 3D.

To achieve above-mentioned performance, a clinically-available hand-held linear-array UT is applied for PA wave detection and image formation in 2D. Meanwhile, a global positioning system records every coordinate of the linear-array UT for each 2D frame, which will be used for 3D image reconstruction from all the acquired 2D frames. Moreover, a dedicated space transformation and reconstruction method for 3D image realization is used in this paper, followed by phantom study and experimental validation *in vivo*.

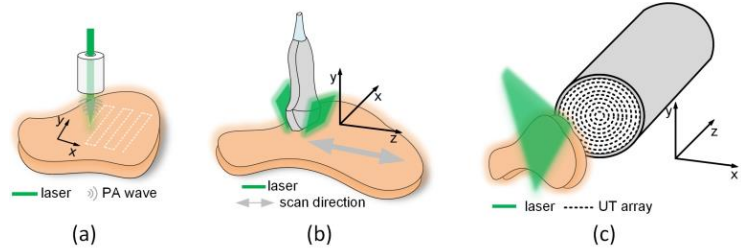

Fig. 1 Conventional 3D photoacoustic imaging implementations. (a) the PAM system with 2D mechanical scanning; (b) 3D PAT realization with UT array mechanical scanning; (c) 3D PAT realization with hemispherical UT array.

## 2. Method

### 2.1 fsPAT System Architecture

The architecture of the proposed fsPAT system is shown in Fig. 2. The laser source of the PAT system is an optical parametric oscillator (OPO) laser (PHOCOUS MOBILE, OPOTEK, USA) with a 10 Hz repetition rate and 690 to 950 nm wavelength tunable rage. The 128-element linear-array UT (Doppler Inc., China) is used for PA signal's detection, whose central frequency is 7.5 MHz with 73% bandwidth. A data acquisition (DAQ) card with 128 channels (PhotoSound, USA) is used for PA data acquisition with sampling rate of 40 MSPS. The 3D global spatial positioning system (G4, Polhemus, USA) is comprised of three parts: system electronics unit (SEU), standard sensor (SS), and electric-magnetic field source (EFS). The EFS generates the magnetic field so that the sensor's trajectory can be precisely tracked as it is moving within the magnetic field. The SEU embedded hardware and software compute the position and orientation of the sensor, and transmit the coordinate data to personal computer (PC). For the proposed fsPAT, the SS is fixed to the ultrasonic probe that tracks its movement during the handheld scanning process. To be more specific, the SS of positioning system is attached in the same plane behind UT probe to avoid the presence of magnetic distortion. Then, the positioning system captures the 6 degrees of freedom (6-DOF) coordinate with 120 Hz sampling rate for each 2D PA image. The acquired 6-DOF coordinates include 3-DOF translation with unit of *mm* $(x, y, z)$ and 3-DOF rotation with unit of degree $(\alpha, \beta, \gamma)$. The whole system is synchronized by proper triggering, controlled by the PC.

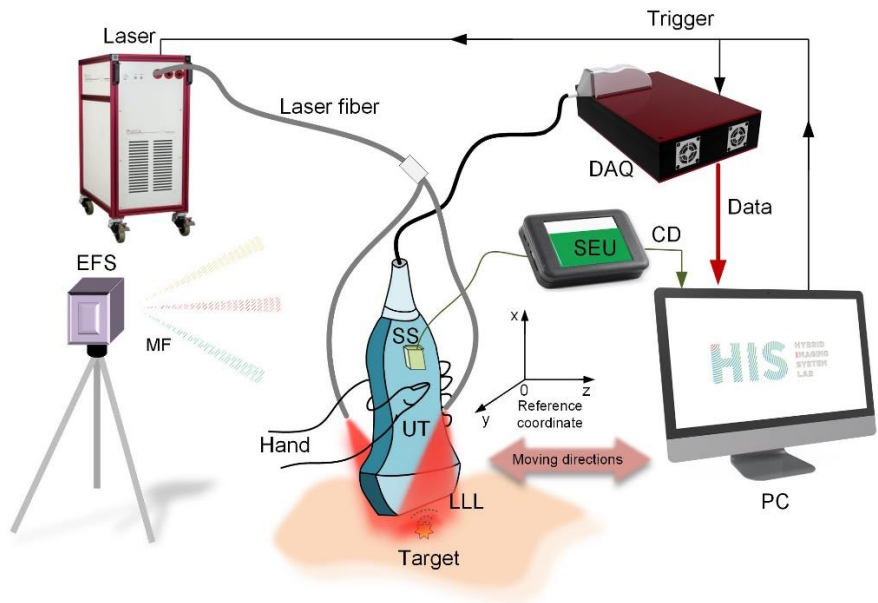

Fig. 2 The setup of the 3D PAT imaging system. DAQ: data acquisition; EFS: electric-magnetic field source; MF: magnetic field; SS: standard sensor; SEU: system electronics unit; CD: coordinate data; UT: ultrasound transducer; LLL: linear laser light; PC: personal computer;

*2.2 fsPAT Working Flow*

The working flow of the proposed fsPAT is presented in Fig. 3. The procedure can be summarized as: (1) Synchronizing signal triggers the laser output to illuminate the imaging target. Following that, PA wave is generated on account of the PA effect; (2) The ultrasound probe captures the PA signals in real time, followed by data acquisition of DAQ system. At the meantime, the standard sensor of the positioning system captures the 6-DOF coordinates of the ultrasound probe; (3) After data acquisition, 2D PA image can be reconstructed by applying universal back-projection algorithms in the computer [20]. (4) When sufficient 2D PA images are collected revealing continuous multiple cross-sections of the imaging target, 3D PA image reconstruction can be performed based on the collected 6-DOF coordinates.

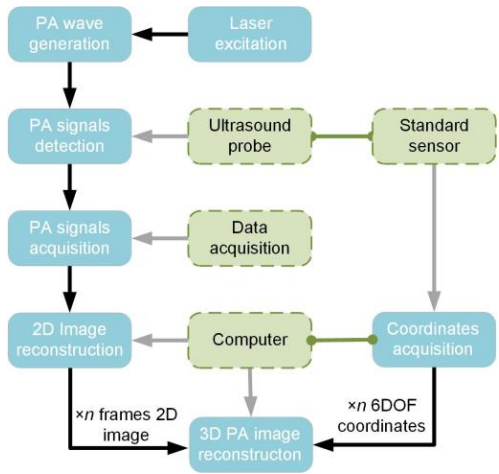

Fig. 3 The free-scan handheld 3D PAT imaging system working flow. DOF: degrees of freedom.

*2.3 3D Space Building and Transformation*

In order to build the geometric relationship for 3D PA image reconstruction, four coordinate spaces in this system are separated: (1) the 2D image space $I$ to express each pixel based on imaging plane; (2) the positioning SS space $P$ to represent pixels according to the installation of sensor; (3) the 3D EFS space $S$ to record 6-DOF position; (4) the volume space $V$ to describe the region of 3D PA imaging.

Once the 2D scan is finished, a constant translation matrix $^{P}T_{I}$ is measured to calibrate the center of probe receiving plane in image space to the sensor space. A matrix $^{S}T_{P}$ is calculated using acquired translation and rotation for the transformation from sensor space to EFS field space. The minimum of transformed coordinate values in *x, y, z* axes are regarded as the origin of volume space. An affine transformation $^{V}T_{S}$ is implemented to move all the acquired location in the EFS space into the view of volume space.

The above processes can be expressed by the following formula:

$$V_x = {}^{V}T_{S} * {}^{S}T_{P} * {}^{P}T_{I} * I_x \tag{1}$$

$$V_x = [x_v \quad y_v \quad z_v \quad 1]^T \tag{2}$$

$$I_x = [u_x \quad v_x \quad 0 \quad 1]^T \tag{3}$$

where $I_x$ and $V_x$ are the homogeneous coordinates of pixel and voxel in the 2D image and volume space, respectively. *T* denotes the operation of transposition.

Finally, by applying Eq. (1), the exact coordinate in volume space of each 2D image can be calculated for the following 3D geometric reconstruction. Fig. 4 shows the visualization result before and after space transformation of a typical PA image scan. Fig. 4(a) is the probe trajectory before transformation. The black rectangle stands for each frame plane of 2D image combined with spatial information in EFS space. Fig. 4(b) is the probe trajectory in volume space after transformation. After all the coordinates transform into volume space, the coordinates can be further used for 3D PA image reconstruction.

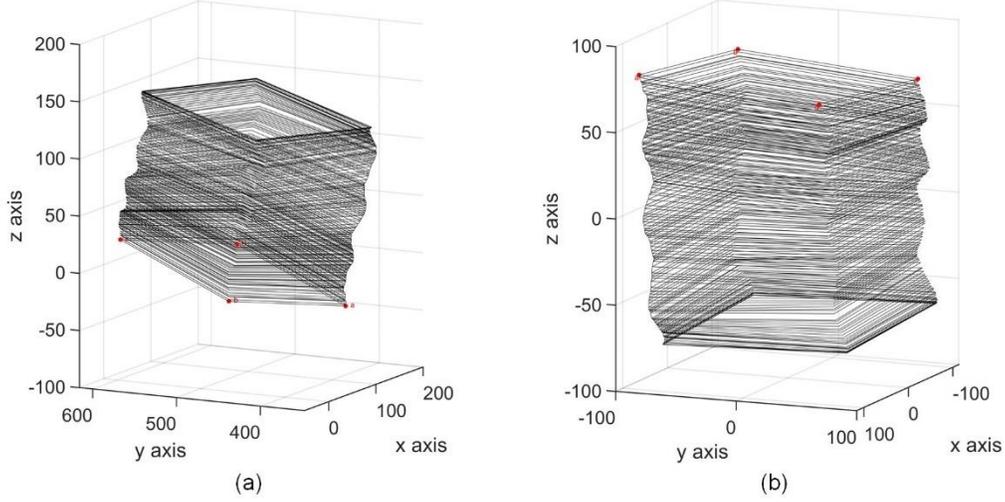

Fig. 4 The visualization of transformation. Each frame plane is described as a rectangle (black). (a) The probe trajectory in EFS space before transformation. (b) The probe trajectory in volume space after transformation.

## *2.4 3D PA Image Reconstruction*

The realization of 3D PA image reconstruction is based on each 2D image acquisition with corresponding spatial position. Fig. 5 shows the pipeline of 3D PA image reconstruction procedure. During the reconstruction process, the fsPAT system will record 128 channels' raw PA data and a 6-DOF spatial coordinate for each 2D PA image, which are then combined for 3D image reconstruction.

To be more specific, a regular grid volume is constructed through the coordinates of all the pixels after transformation by Eq. (1). The voxel size of this volume is set as 0.2 *mm* × 0.2 *mm* × 0.6 *mm (x* direction, *y* direction and *z* direction*)*. Then, 2D PA images labeled with position information are fused into the grid volume for recovering the mapping connection between pixels and voxels. Here, the reconstruction algorithm applied in this paper is a Forward mapping algorithm with acceleration, the Pixel Nearest Neighbor based Fast-Dot Projection (FDP-PNN) [21, 22].

The mapping result of a pixel point $I_0$ in the 2D image can be acquired by the nearest voxel in 3D volume around it:

$$V_0 = \min({}^V T_S * {}^S T_P * {}^P T_I * I_0) \tag{4}$$

where $V_0$ is a 1×3 coordinate vector of the voxel, which is mapped by $I_0$. The corresponding voxels of other pixels in the same 2D image can be computed as following:

$$V_i = d_w * \Delta w + d_h * \Delta h + V_o \tag{5}$$

where, for a 2D image, $d_w$ and $d_h$ are the normalized direction vector in both axes of width and height. $\Delta w$ and $\Delta h$ are the difference between the index of each pixel and $I_0$. Lastly, the intensity of mapped voxel is assigned by the value of its matching pixel.

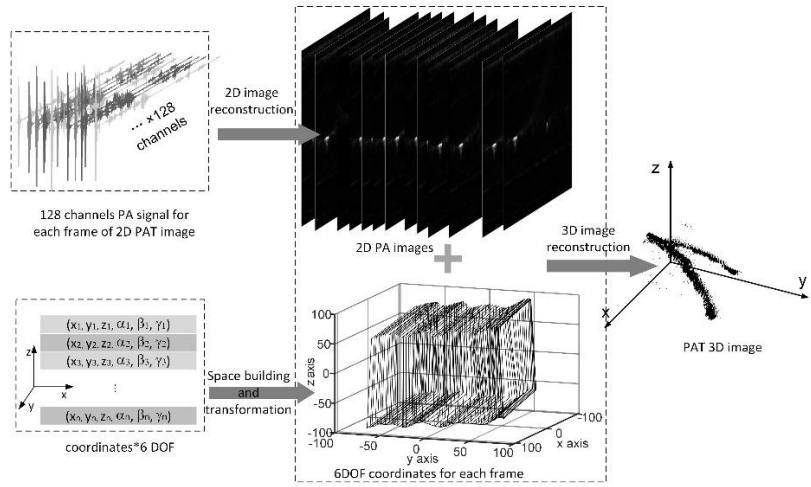

Fig. 5 The pipeline of fsPAT 3D image reconstruction procedure.

## 3. Experimental Results

### 3.1 Phantom Study

The phantom of human hair embedded in agar gel is used for experimental feasibility study. Fig. 6(a) shows the photograph of the phantom, where the hair is about 7~10 cm long. The imaging frame rate is limited by the laser repetition rate at 10 Hz. The whole scanning time consumption is about half-minute, capturing 300 frames of the 2D PA images in total. Each 2D PA image is reconstructed with $256 \times 256$ pixel size.

After the space building and transformation, and by applying the 3D image reconstruction algorithms mentioned above, the 3D PA imaging of the hair phantom is shown in Fig. 6(b)-(d) corresponding to the top view, lateral view and front view, respectively. The 3D imaging result reveals the structure of the phantom clearly with high contrast to the background. The black arrows in Fig. 6(a) mark several positions of the phantom, and their corresponding positions in the PA images are pointed by the white arrows in Fig. 6(b)-(d).

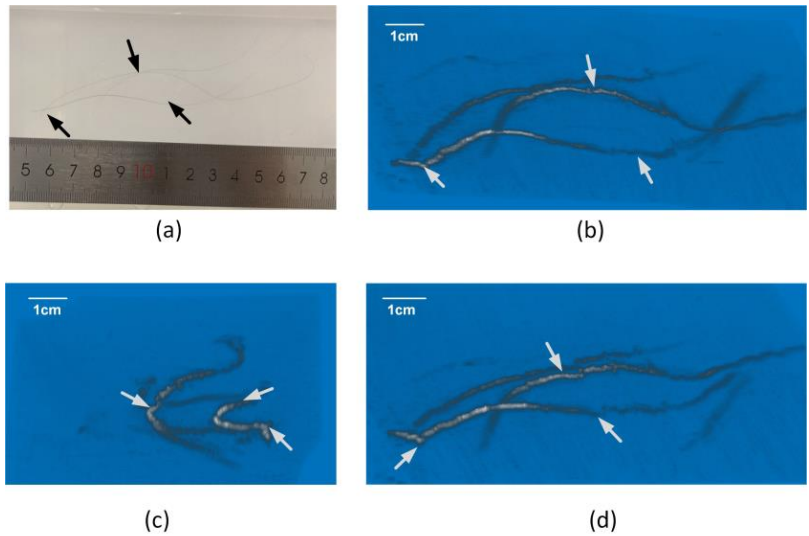

Fig. 6 The phantom experimental result. (a) The photograph of the phantom (human hair embedding in agar gel); (b)-(d) the reconstructed 3D PA imaging of the hair phantom corresponding to the top view, lateral view and front view, respectively.

## 3.2 Quantitative Analysis

A pencil leads made phantom is used for quantitative analysis for the performance evaluation of the proposed system. Three pencil leads are placed parallel with different depths in the agar gel, and the other three are standing upright in agar gel. A total of nine physical distances $d_1 \sim d_9$ are measured using a caliper indicated in Fig. 7(a)-(b), including three lengths of leads *($d_1$, $d_2$ and $d_3$)* and six interval distances *($d_4 \sim d_9$)*. Fig. 7(c) shows the reconstructed 3D structure of the parallel placed three pencil leads. The projection images on the top view of the 3-D reconstruction result for the leads phantom are shown in Fig. 7 (d)-(e). The same nine distances are measured by selecting the pixels on the projection image, calculating the Euclidean distances between these pixels with multiplication by voxel size.

Table 1 presents the measurement results of these distances. The absolute error and relative error are applied to analyze the accuracy between the physical distance of the phantom and the PA imaging result. The mean and standard deviation of these two errors for all the distances are 0.22*mm*±0.21*mm* and 2.11%±2.18%, respectively.

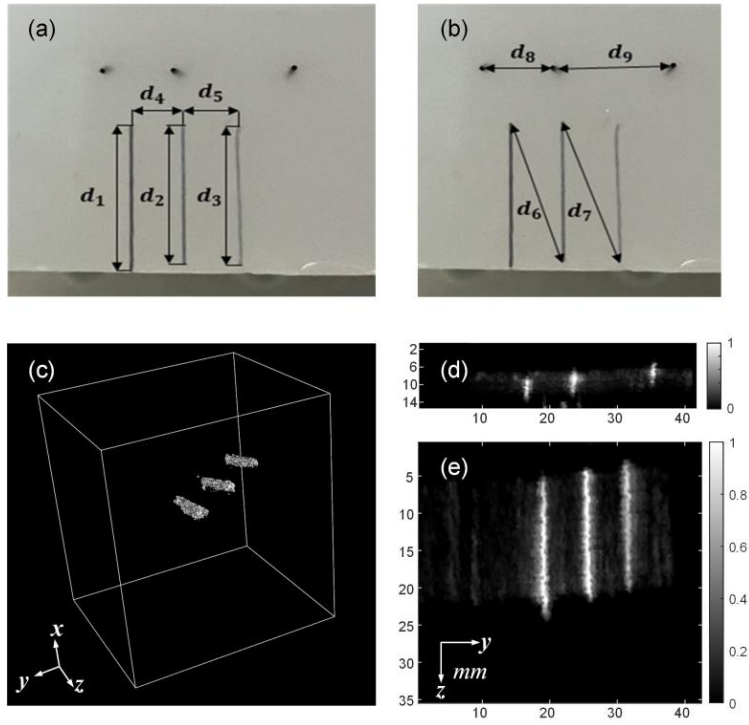

Fig. 7 The photograph of pencil leads phantom and experimental results. (a) The photograph of the pencil leads phantom (illustration of three distances of leads' length $(d_1, d_2$ and $d_3)$ and two distances of horizontal interval $(d_4$ and $d_5)$); (b) The pencil leads phantom with illustration of two distances of oblique interval $(d_6$ and $d_7)$ and two distances of horizontal interval $(d_8$ and $d_9)$; (c) the reconstructed 3D structure of the parallel placed three pencil leads; (d) Projection image of three standing leads on top view of 3-D imaging volume. (e) Projection image of three paralleled leads on top view of 3-D imaging volume.

**Table 1. The Analysis of Measurement Result**

| Distance (*mm*) | Physical distance | Reconstructed distance | Absolute error | Relative error |
|---|---|---|---|---|
| $d_1$ | 18.70 | 18.91 | 0.21 | 1.10% |
| $d_2$ | 18.17 | 18.04 | 0.13 | 0.72% |
| $d_3$ | 18.22 | 18.07 | 0.15 | 0.82% |
| $d_4$ | 5.70 | 5.80 | 0.10 | 1.75% |
| $d_5$ | 6.80 | 6.40 | 0.40 | 5.88% |
| $d_6$ | 20.45 | 20.53 | 0.08 | 0.37% |
| $d_7$ | 20.22 | 20.24 | 0.02 | 0.11% |
| $d_8$ | 7.40 | 7.60 | 0.20 | 2.74% |
| $d_9$ | 12.90 | 13.61 | 0.71 | 5.53% |
| | | Mean Error | 0.22 | 2.11% |
| | | Standard deviation | 0.21 | 2.18% |

*3.3 In-vivo Human Study*

The 3D PA imaging of human left-wrist vascular is also performed by the proposed fsPAT. To get better penetration of light, the wavelength is changed to 1064 nm with another high-power laser (LPS-1064-L, CNI, China) with 120 $mJ$ pulse energy at the fiber output with 10 Hz repetition rate. The light is transmitted by an optical fiber and illuminated on the surface of skin with illumination area of about 8 cm$^2$. The fluence is less than 20 $mJ/cm^2$ at the skin surface within the safety limit [23]. In this experiment, the hand wrist and forearm are immersed in water with a constant temperature of 28 °C. The photograph of the experimental operation is shown in Fig. 8(a).

There are totally 240 frames captured in the ROI from the middle of the forearm to wrist with scanning range of about 50 mm. The reconstructed 3D PA image of vessels is shown in Fig. 8(b)-(d). Fig. 8(b) shows the top view of the reconstructed 3D structure of the blood vessel, where the outline of the blood vessel is obvious with striking contrast to the background. Fig. 8(c) shows the lateral view of the blood vessel, where the depth of the blood vessel is revealed in this view. Fig. 8(d) is the left top view of the reconstructed 3D PA image. The arrows labelled ① to ④ in Fig. 8(b)-(d) correspond to the structures of *Ulnar vein*, a part of *Superficial Palmar venous arch*, *Median antebrachial vein* and *Radial vein* in anatomy [24]. Generally, the human wrist imaging results show clear outline of the blood vessel with complete structure, which well proved its feasibility for clinical applications in the near future.

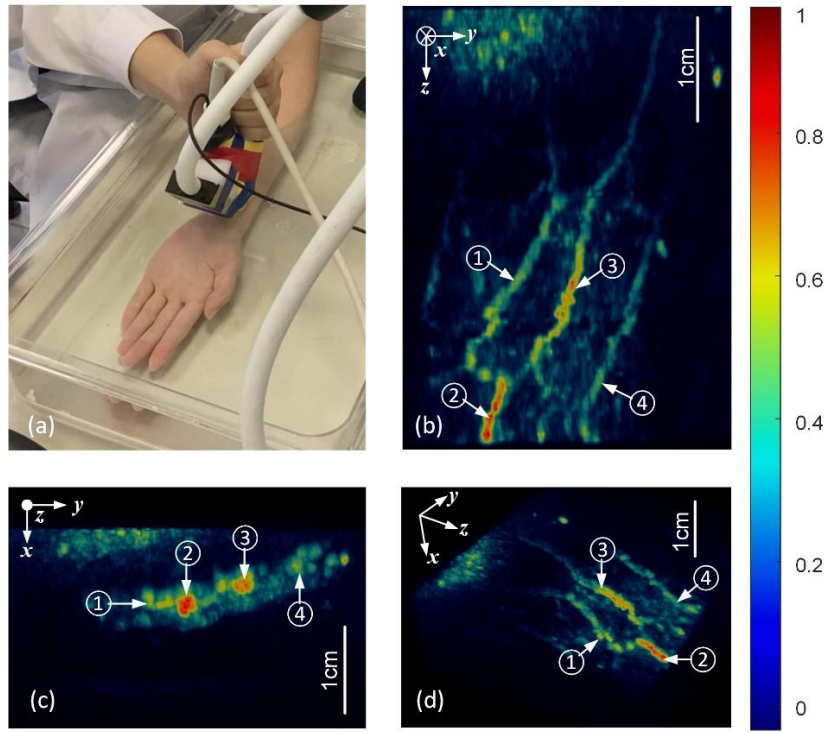

Fig. 8 In-vivo experimental results of human wrist. (a) the photograph of the experimental operation with the proposed fsPAT. (b)-(d) the 3D PA imaging of human wrist blood vessel in top view, lateral view and left top view. ①: Ulnar vein; ②: Superficial Palmar venous arch; ③: Median antebrachial vein; ④: Radial vein.

**4. Discussion**

In this paper, we proposed fsPAT system, which contains 128-channel UT probe for 2D PA image acquisition, and a global positioning system for spatial coordinates collection of the 2D image. The fsPAT system realized 3D PAT imaging with free-scanning handheld probe. The imaging ROI of the system is significantly improved with handheld scanning. The global positioning system captures the spatial information with 120 Hz, which records the coordinates of 2D image fast enough. The 2D imaging is with 10 Hz rate and 38.4 mm imaging width. Therefore, the fsPAT system's ROI has cross-sectional width of 38.4 mm, and the length of the ROI is based on the handheld probe scanning region. With 10 Hz laser repetition rate and 0.1~0.4 mm scanning interval, the fsPAT can realize 60~240 mm per minute scanning speed for 3D imaging. Moreover, the scanning speed can be further improved by applying the laser source with higher repetition rate.

In the phantom study of human hair embedded in agar gel, the feasibility of the proposed fsPAT system for 3D photoacoustic imaging is verified. The fsPAT realized 7~10 cm length of large ROI imaging by handheld free-scanning, which is flexible for clinical applications. The imaging results in Fig. 6(b)-(d) reveal the 3D structure and size of the phantom with strong contrast to the background. Furthermore, the quantitative analysis of the fsPAT for 3D imaging is verified by a pencil leads made phantom. There are nine Euclidean distances measured in the phantom as shown Fig. 7(a)-(b), and we calculated the corresponding distances in the reconstructed imaging results. Table 1 shows the measurement results, where the utmost relative error is 5.88%. The mean absolute error of the nine measurements is 0.22 mm with 2.11% relative error, which shows satisfactory accuracy of the reconstructed 3D structure.

The in-vivo imaging results of human wrist are shown in Fig. 8, showing that the fsPAT system with handheld free-scanning realized a large ROI 3D imaging, where the reconstructed 3D image shows the size and shape of the blood vessel. The *in-vivo* imaging results of the human wrist blood vessel show high potential of the fsPAT system for clinical 3D photoacoustic imaging applications.

However, the proposed fsPAT also has some limitations and need to be further optimized: 1) the 3D imaging speed for large ROI is limited by the 2D image realization, e.g. the repetition rate of the laser source; 2) the cross-sectional width of the 3D image is limited by the UT probe receiving width; 3) the 2D image distortion induce the challenge for exactly 3D reconstruction. To further advance the fsPAT for 3D imaging, the reconstruction algorithms from raw PA signals directly to 3D image will be developed to achieve isotropic-resolution 3D PA imaging.

## 5. Conclusion

The integration of the global positioning system and PAT system enables the proposed fsPAT to realize large ROI of 3D imaging by freely handheld scanning, which greatly improves the flexibility of the 3D PA imaging compared with the conventional 3D PA imaging methods. In this paper, both phantom study and *in-vivo* imaging are conducted that verified the feasibility and performance of the proposed system. Furthermore, in-vivo human wrist blood vessel imaging shows potential for clinical applications such as peripheral vessel visualization, and breast cancer detection.

In the future work, fsPAT can be further optimized by employing laser with high repetition rate for fast imaging scanning, and multi-wavelength for functional 3D imaging e.g. blood oxygen saturation quantification and imaging. In addition, the fsPAT can be also combined with ultrasound imaging to visual more details of vessels.